\RequirePackage{lineno}
\documentclass[aps,twocolumn,showpacs,byrevtex,reprint]{revtex4-1}

\usepackage{graphicx}
\usepackage{dcolumn}
\usepackage{bm}
\usepackage{rotating}
\usepackage{epstopdf}
\usepackage{color}
\usepackage{verbatim} 
\usepackage{multirow}
\usepackage[abs]{overpic}
\usepackage{amsmath}

\newcommand{\PreserveBackslash}[1]{\let\temp=\\#1\let\\=\temp}
\newcolumntype{C}[1]{>{\PreserveBackslash\centering}p{#1}}
\newcolumntype{R}[1]{>{\PreserveBackslash\raggedleft}p{#1}}
\newcolumntype{L}[1]{>{\PreserveBackslash\raggedright}p{#1}}

\newcommand{\EE}{e^+e^-}

\newcommand{\too}{\rightarrow}

\begin{document}
\graphicspath{{figure/}}
\DeclareGraphicsExtensions{.eps,.png,.ps}

\title{\quad\\[0.0cm] \boldmath Search for vector charmonium(-like) states in the $\EE \too \eta J/\psi$ line shape}

\author{Jielei Zhang}
\email{zhangjielei@ihep.ac.cn}
\author{Rumin Wang}
\affiliation{College of Physics and Electronic Engineering, Xinyang Normal University, Xinyang 464000, People's Republic of China}

\begin{abstract}
The cross section of $\EE \too \eta J/\psi$ has been measured by BESIII and Belle experiments. Fit to the $\EE \too \eta J/\psi$ line shape, three resonant structures are evident. The parameters for the three resonant structures are $M_{1}=(3980\pm17\pm7)$ MeV/$c^{2}$, $\Gamma_{1}=(104\pm32\pm13)$ MeV; $M_{2}=(4219\pm5\pm4)$ MeV/$c^{2}$, $\Gamma_{2}=(63\pm9\pm3)$ MeV; $M_{3}=(4401\pm12\pm4)$ MeV/$c^{2}$, $\Gamma_{3}=(49\pm19\pm4)$ MeV, where the first uncertainties are statistical and the second systematic. We attribute the three structures to $\psi(4040)$, $Y(4220)$ and $\psi(4415)$ states. The branching fractions $\mathcal{B}(\psi(4040) \too \eta J/\psi)$ and $\mathcal{B}(\psi(4415) \too \eta J/\psi)$ are given. If $Y(4220)$ is taken as $\psi(4S)$ state, the branching fraction $\mathcal{B}(\psi(4S) \too \eta J/\psi)$ is also given. Combining all $Y(4220)$ parameters obtained from different decays, we give average parameters for $Y(4220)$, which are $M_{Y(4220)}=(4220.8\pm2.4)$ MeV/$c^{2}$, $\Gamma_{Y(4220)}=(54.8\pm3.3)$ MeV.
\end{abstract}

\maketitle

The potential model works well in describing the heavy quarkonia states~\cite{potential}, especially for the charmonium states below the open-charm threshold. However, above this threshold, there are still many predicted states have not been observed yet. In recent years, charmonium physics has gained renewed strong interest from both the theoretical and the experimental side, due to the observation of charmonium-like states, such as $Y(4260)$~\cite{Y4260}, $Y(4360)$~\cite{Y4360} and $Y(4660)$~\cite{Y4660}. These $Y$-states are above the open-charm threshold and do not fit in the conventional charmonium spectroscopy, so they could be exotic states that lie outside the quark model~\cite{exotic, exotic2, exotic3}. The $1^{--}$ $Y$-states are all observed in $\pi^{+}\pi^{-}J/\psi$ or $\pi^{+}\pi^{-}\psi(3686)$, while recently, one state (called $Y(4220)$) is observed in $\EE \too \omega \chi_{c0}$~\cite{omegachic2}, and two states (called $Y(4220)$ and $Y(4390)$) are observed in $\EE \too \pi^{+}\pi^{-}h_{c}$~\cite{pipihc, pipihc2}. It indicates that the $Y$-states also can be searched by other charmonium transition decays. Among them, the cross section for $\EE \too \eta J/\psi$ is relative large, so we can search for $Y$-states in $\eta J/\psi$ line shape. The study of these $1^{--}$ $Y$-states is very helpful to clarify the missing predicted charmonium states in potential model. In all $Y$-states, maybe some are conventional charmonium. So it is important to confirm that which $Y$-states are charmonium and which $Y$-states are exotic states.

\begin{figure}[htbp]
\begin{center}
\includegraphics[width=0.43\textwidth]{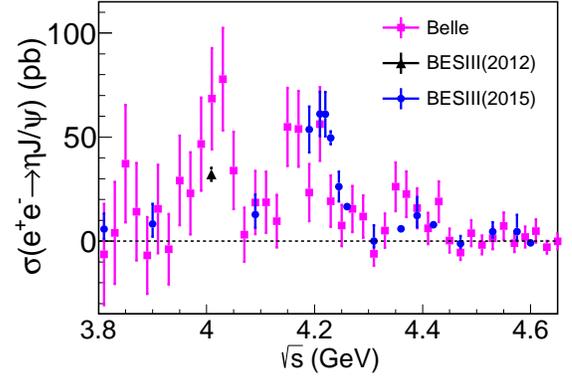}
\caption{Cross section of $\EE \too \eta J/\psi$ from BESIII and Belle experiments.}
\label{fig:crosssection}
\end{center}
\end{figure}

Recently, the process $\EE \too \eta J/\psi$ has been measured by BESIII~\cite{bes1,bes2} and Belle~\cite{belle} experiments. In Ref.~\cite{belle}, authors claim $\eta J/\psi$ is from resonances $\psi(4040)$ and $\psi(4160)$. Figure~\ref{fig:crosssection} shows the cross sections from the two experiments for the center-of-mass energy from 3.80 to 4.65 GeV, and they are consistent with each other within error. The cross section of $\EE \too \eta J/\psi$ is of the same order of magnitude as those of the $\EE \too \pi^{+}\pi^{-} J/\psi$~\cite{pipijpsi} or $\EE \too \pi^{+}\pi^{-} \psi(3686)$~\cite{pipipsip}, but with a different line shape. It indicates there is large coupling between $Y$-states and $\eta J/\psi$. So we try to use BESIII and Belle measurements to extract the resonant structures parameters in $\EE \too \eta J/\psi$. Currently, between 3.80 GeV and 4.65 GeV, the all observed vector charmonium(-like) states are $\psi(4040)$, $\psi(4160)$,  $Y(4220)$, $Y(4360)$, $Y(4390)$ and $\psi(4415)$. In this paper, We try to search for these vector charmonium(-like) states in the $\EE \too \eta J/\psi$ line shape.

We use a least $\chi^{2}$ method to fit the cross section. From Fig.~\ref{fig:crosssection}, we can see there are three obvious structures around 4, 4.2 and 4.4 GeV in the line shape of $\EE \too \eta J/\psi$. Assuming that $\eta J/\psi$ comes from three resonances, we fit the cross section with coherent sum of three constant width relativistic Breit-Wigner (BW) function (model 1, $BW_1+BW_2+BW_3$); that is,
\begin{equation}
\begin{aligned}
\sigma(\sqrt{s})= & |BW_{1}(\sqrt{s})\sqrt{\frac{PS^{3}(\sqrt{s})}{PS^{3}(M_{1})}}+BW_{2}(\sqrt{s}) \\
&¡¡\sqrt{\frac{PS^{3}(\sqrt{s})}{PS^{3}(M_{2})}}e^{i\phi_{1}} +BW_{3}(\sqrt{s})\sqrt{\frac{PS^{3}(\sqrt{s})}{PS^{3}(M_{3})}}e^{i\phi_{2}}|^{2},
\end{aligned}
\end{equation}
where $PS(\sqrt{s})=p/\sqrt{s}$ is the 2-body phase space factor, where $p$ is the $\eta$ or $J/\psi$ momentum in the $\EE$ center-of-mass frame, $\phi_{1}$ and $\phi_{2}$ are relative phases, $BW(\sqrt{s})=\frac{\sqrt{12\pi\Gamma_{ee}\mathcal{B}(\eta J/\psi)\Gamma}}{s-M^{2}+iM\Gamma}$, is the BW function for a vector state, with mass $M$, total width $\Gamma$, electron partial width $\Gamma_{ee}$, and the branching fraction to $\eta J/\psi$, $\mathcal{B}(\eta J/\psi)$. From the fit, the $\Gamma_{ee}$ and $\mathcal{B}(\eta J/\psi)$ can not be obtained separately, we can only extract the product $\Gamma_{ee}\mathcal{B}(\eta J/\psi)$.

\begin{figure}[htbp]
\begin{center}
\begin{overpic}[width=0.43\textwidth]{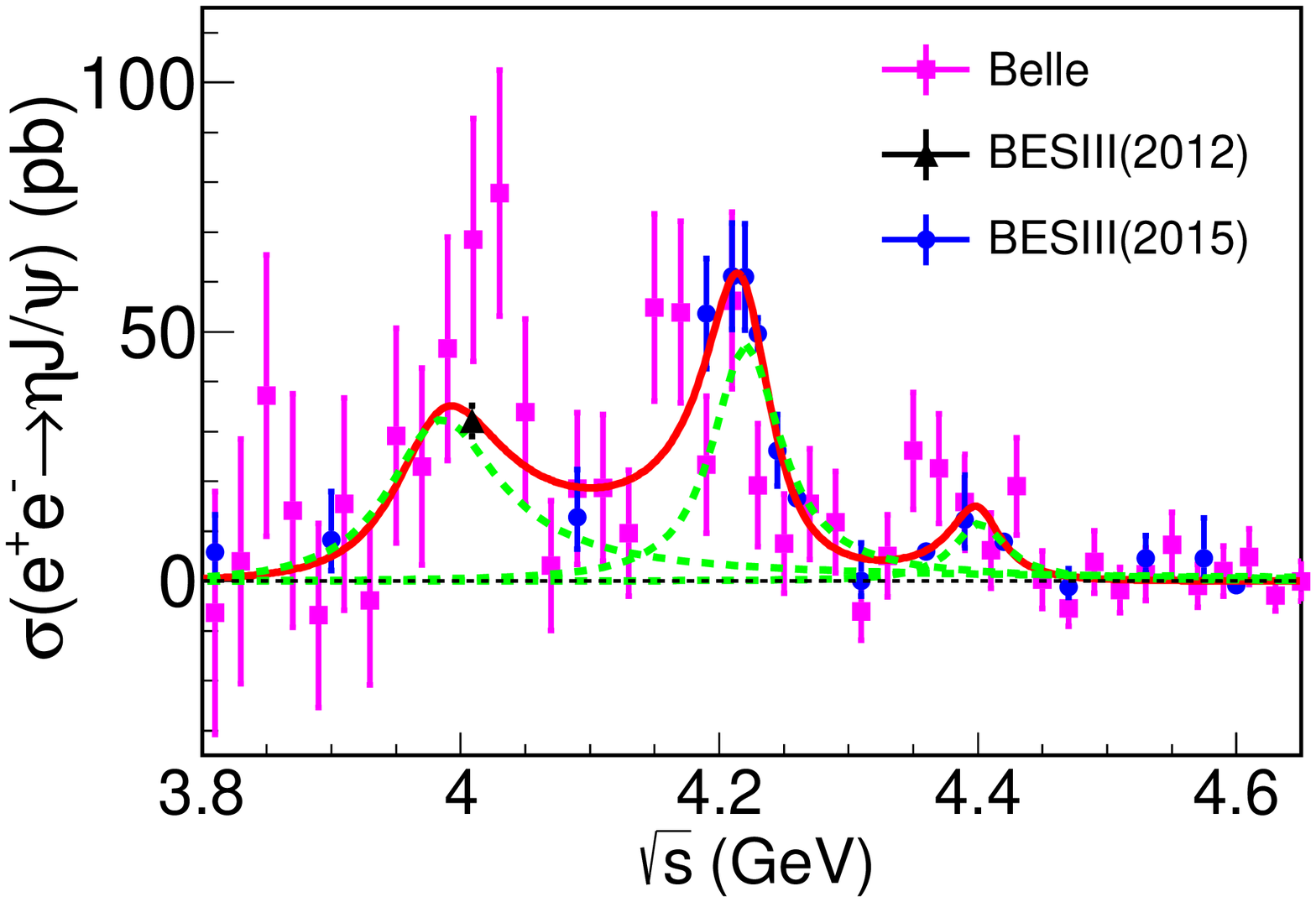}
\put(48,123){\large (a)}
\end{overpic}
\begin{overpic}[width=0.43\textwidth]{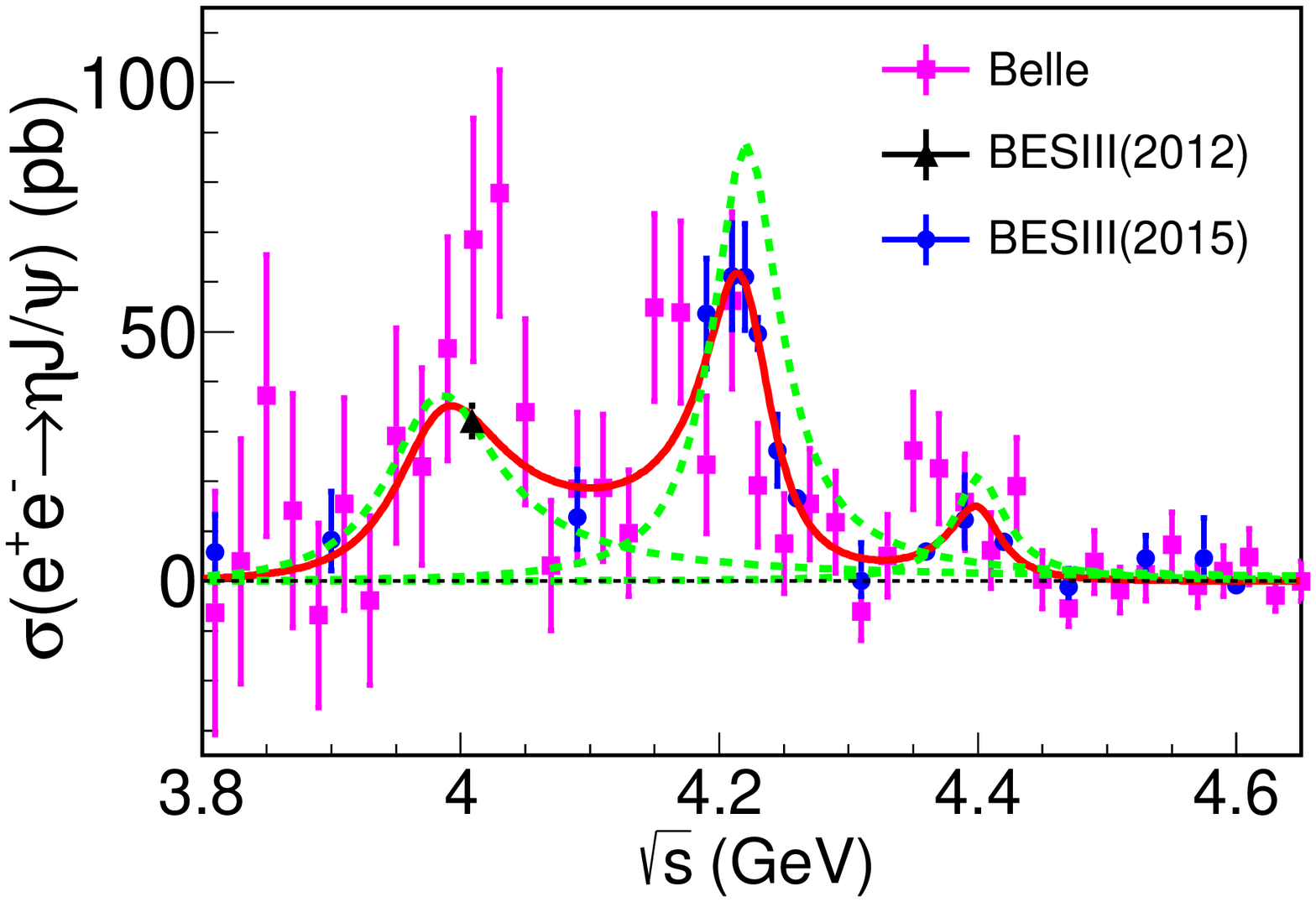}
\put(48,123){\large (b)}
\end{overpic}
\caption{The fits to the cross section of $\EE \too \eta J/\psi$ from BESIII and Belle experiments, two solutions are found and are shown in plots (a) and (b). The solid red curves show the best fits, and the dashed green ones are individual components.}
\label{fig:fit1}
\end{center}
\end{figure}

The $\chi^{2}$ is minimized to obtain the best estimation of the resonant parameters, and the statistical uncertainties are obtained when $\chi^{2}$ value change 1 compared with the minimum. Figure~\ref{fig:fit1} shows the fit results. Two solutions are found with the same fit quality. The fits indicate the existence of three resonances (called $Y_{1}$, $Y_{2}$, $Y_{3}$) with mass and width are $M_{1}=(3980\pm17)$ MeV/$c^{2}$, $\Gamma_{1}=(104\pm32)$ MeV; $M_{2}=(4219\pm5)$ MeV/$c^{2}$, $\Gamma_{2}=(63\pm9)$ MeV; $M_{3}=(4401\pm12)$ MeV/$c^{2}$, $\Gamma_{3}=(49\pm19)$ MeV, and the goodness of the fit is $\chi^{2}/ndf=51.8/50$, corresponding to a confidence level of $40\%$, where $ndf$ is the number of degrees of freedom. The all fitted parameters of the cross section of $\EE \too \eta J/\psi$ are listed in Table~\ref{tab:fitresult1}.

\begin{table*}[htbp]
\begin{center}
\caption{ The fitted parameters of the cross section of $\EE \too \eta J/\psi$. The first uncertainties are statistical, and the second systematic. }
\label{tab:fitresult1}
\begin{tabular}{ccc}
  \hline
  \hline
  \qquad \qquad Parameter \qquad \qquad \qquad & \qquad \qquad SolutionI \qquad \qquad \qquad & \qquad \qquad SolutionII \qquad \qquad \qquad  \\
  \hline
  $M_{1}$ (MeV/$c^{2}$) & \multicolumn{2}{c}{$3980\pm17\pm7$} \\
  $\Gamma_{1}$ (MeV) & \multicolumn{2}{c}{$104\pm32\pm13$} \\
  $\Gamma^{Y_{1}}_{ee}\mathcal{B}(Y_{1} \too \eta J/\psi)$ (eV) & $3.6\pm0.9\pm0.3$ & $4.1\pm1.2\pm0.4$ \\
  $M_{2}$ (MeV/$c^{2}$) & \multicolumn{2}{c}{$4219\pm5\pm4$} \\
  $\Gamma_{2}$ (MeV) & \multicolumn{2}{c}{$63\pm9\pm3$} \\
  $\Gamma^{Y_{2}}_{ee}\mathcal{B}(Y_{2} \too \eta J/\psi)$ (eV) & $3.6\pm1.1\pm0.3$ & $6.7\pm1.3\pm0.4$ \\
  $M_{3}$ (MeV/$c^{2}$) & \multicolumn{2}{c}{$4401\pm12\pm4$} \\
  $\Gamma_{3}$ (MeV) & \multicolumn{2}{c}{$49\pm19\pm4$} \\
  $\Gamma^{Y_{3}}_{ee}\mathcal{B}(Y_{3} \too \eta J/\psi)$ (eV) & $0.7\pm0.3\pm0.2$ & $1.4\pm0.7\pm0.2$ \\
  $\phi_{1}$ & $2.76\pm0.53\pm0.19$ & $-2.64\pm0.43\pm0.18$ \\
  $\phi_{2}$ & $-2.34\pm1.10\pm0.25$ & $1.88\pm1.03\pm0.24$ \\
  \hline
  \hline
\end{tabular}
\end{center}
\end{table*}

There is only one $1^{--}$ charmonium state around 4.2 GeV, which is $\psi(4160)$. It is interesting to check whether the second structure in $\EE \too \eta J/\psi$ is from $\psi(4160)$, or whether there is contribution from $\psi(4160)$. If we fix $Y_{2}$ state's parameters to the mass and width of $\psi(4160)$ (model 2, $BW_1+\psi(4160)+BW_3$)~\cite{pdg}, the goodness of the fit is $\chi^{2}/ndf=101.7/52$, corresponding to a confidence level of $5\times10^{-5}$. We also try to add $\psi(4160)$ resonance to fit the cross section (model 3, $BW_1+BW_2+\psi(4160)+BW_3$), where the mass and width of $\psi(4160)$ are fixed at the world average values~\cite{pdg} in the fit. The goodness of the fit is $\chi^{2}/ndf=51.0/48$, corresponding to a confidence level of $36\%$.

The systematic uncertainties will also influence on the goodness of fit. The systematic uncertainties includes the uncertainty of the center-of-mass energy determination, parametrization of the BW function, the cross section measurement and the uncertainty of $\psi(4160)$'s mass and width. Since the uncertainty of the measured beam energy is about 0.8 MeV at BESIII, so the beam energy $\sqrt{s}$ is varied within 0.8 MeV in the fit. To estimate the uncertainty from parametrization of BW function, the width $\Gamma$ of BW function is set to be the mass dependent width $\Gamma=\Gamma^{0}\frac{PS(\sqrt{s})}{PS(M)}$ in the fit, where $\Gamma^{0}$ is the width of the resonance. The uncertainty of the cross section measurements will affect the goodness of fit, we vary the cross section values within the systematic uncertainty in the fit. To estimate the uncertainty from the uncertainty of $\psi(4160)$'s mass and width, we vary $\psi(4160)$'s mass and width within the uncertainty to refit. The results for different situations are also listed in Table~\ref{tab:check}.
\begin{table*}[htbp]
\begin{center}
\caption{ The goodness of fit values for different systematic uncertainty terms in different models, which $ndf$ has not been divided. The $ndf$ are 50, 52 and 48 for models $BW_1+BW_2+BW_3$, $BW_1+\psi(4160)+BW_3$ and $BW_1+BW_2+\psi(4160)+BW_3$, respectively. The first value in brackets is the minimum value for different situations, and the second value is the maximum. The column ``Range" is the range of goodness of fit value for all different systematic uncertainty situations, and the column ``C.L." is the range of confidence level corresponding to the value in column ``Range". }
\label{tab:check}
\begin{tabular}{ccccccc}
  \hline
  \hline
   Model & Beam energy & BW function & Cross section & $\psi(4160)$ parameters & Range  & C.L. \\
  \hline
  $BW_1+BW_2+BW_3$  & $(51.2,52.4)$ & $51.8$ & $(50.2,55.5)$ & - & $(49.7, 56.3)$ & $(25\%,49\%)$ \\
  $BW_1+\psi(4160)+BW_3$ & $(99.1,102.2)$ & $111.1$ & $(97.5,109.7)$ & $(86.2,120.1)$ & $(82.4, 139.5)$ & $(9\times10^{-10},0.5\%)$ \\
  $BW_1+BW_2+\psi(4160)+BW_3$ & $(50.3,51.7)$ & $50.9$ & $(49.3,54.7)$ & $(50.5,51.3)$ & $(48.0, 55.9)$ & $(20\%,47\%)$ \\
  \hline
  \hline
\end{tabular}
\end{center}
\end{table*}

From the Table~\ref{tab:check}, we notice that the model 2's confidence level is very bad, while model 1 and model 3's are all good. To get the significance of $BW_2$, we choose model 2 as zero hypothesis, and choose model 3 as alternative hypothesis. As when model 2 is at the minimum (maximum) goodness of fit value, model 3 is also at the minimum (maximum) goodness of fit value. So the statistical significance of $BW_2$ is $(5.0\sigma, 8.4\sigma)$, comparing the $\chi^{2}$s change and taking into the change of the number of degree of freedom. To get the significance of $\psi(4160)$, we choose model 1 as zero hypothesis, and choose model 3 as alternative hypothesis. The statistical significance of $\psi(4160)$ is $(0.2\sigma, 0.8\sigma)$, it means that $\psi(4160)$ resonance is not significance. According to the above analysis, to describe the second structure in the fit, we only need one resonance $BW_2$, the resonance $\psi(4160)$ is not necessary.

To check the statistical significance of the third structure, we choose $BW_1+BW_2$ as zero hypothesis, then get the statistical significance of the third resonance is $5.6\sigma$, comparing the $\chi^{2}$s change and taking into the change of the number of degree of freedom. So to describe the $\EE \too \eta J/\psi$ line shape very well, three resonances is required, the fourth resonance is not necessary based on the current data. The model 1 ($BW_1+BW_2+BW_3$) can describe the line shape very well.

The systematic uncertainties on the resonant parameters are mainly from the uncertainty of the center-of-mass energy determination, parametrization of the BW function, and the cross section measurement. The details have been described above. By assuming all these sources of systematic uncertainties are independent, we add them in quadrature.

\begin{figure}[htbp]
\begin{center}
\begin{overpic}[width=0.46\textwidth]{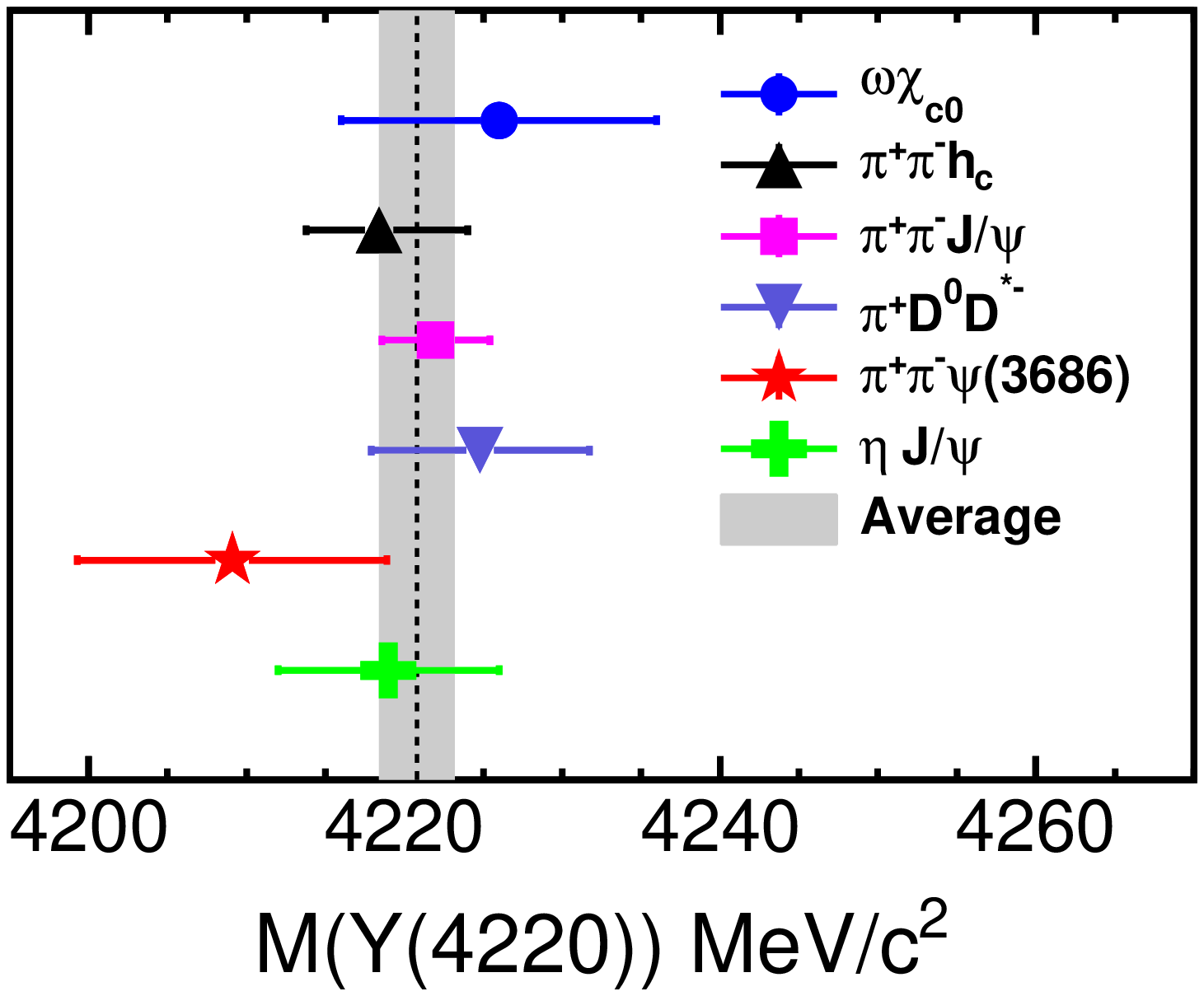}
\put(192,127){\large (a)}
\end{overpic}
\begin{overpic}[width=0.46\textwidth]{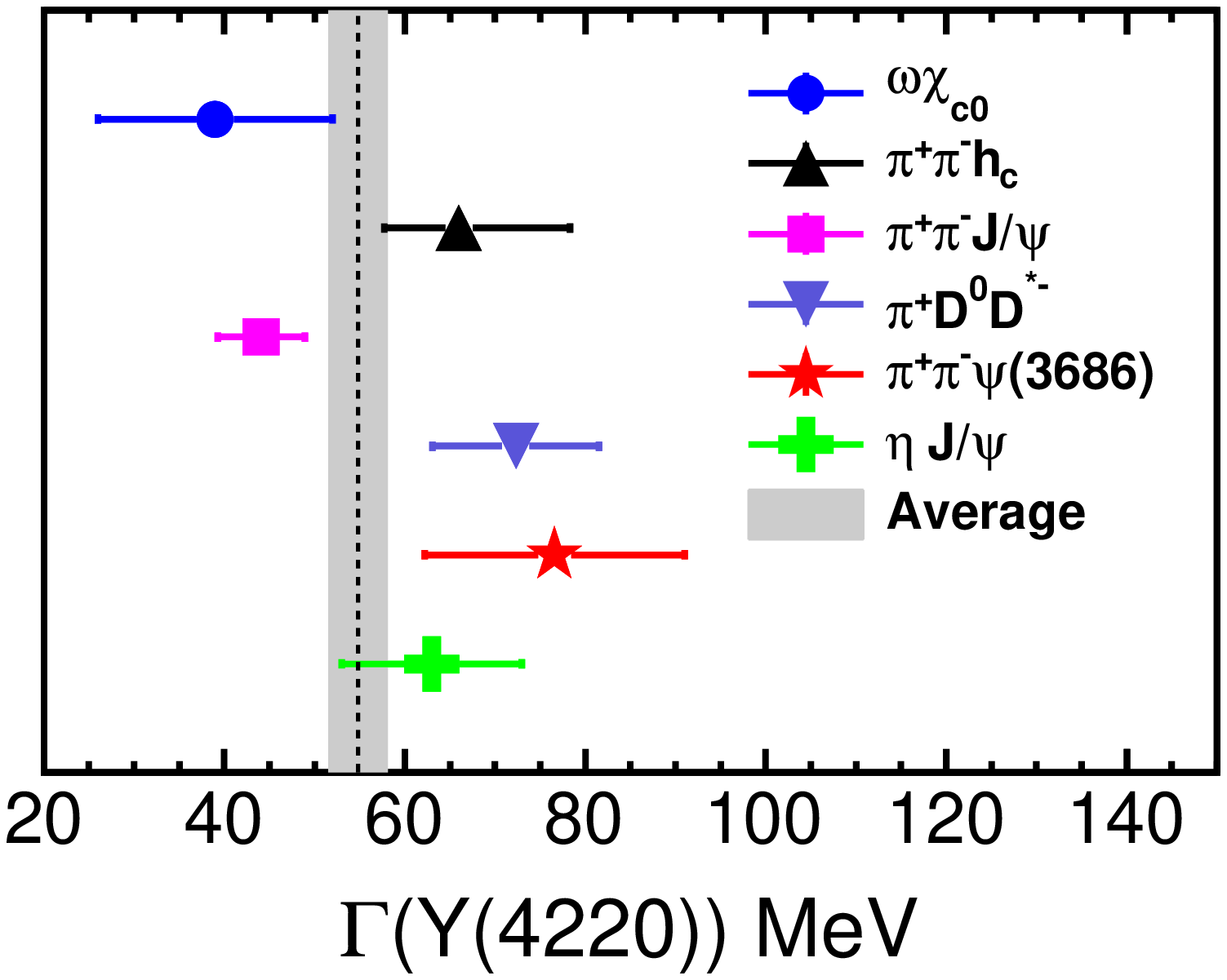}
\put(192,127){\large (b)}
\end{overpic}
\caption{The fits to the mass (plot (a)) and width (plot (b)) of $Y(4220)$ from different decay modes. The grey hatched regions are the average values with uncertainty from fit results.}
\label{fig:fitY4220}
\end{center}
\end{figure}

From above fit results using model 1, we notice that $Y_{1}$'s parameters are close to $\psi(4040)$~\cite{pdg}, the differences are about $3\sigma$ and less than $1\sigma$ for mass and width, and $Y_{3}$'s parameters are close to $\psi(4415)$~\cite{pdg}, the differences are less than $2\sigma$ and less than $1\sigma$ for mass and width, so here we attribute the $Y_{1}$ and $Y_{3}$ states to $\psi(4040)$ and $\psi(4415)$ states. If we take $\Gamma(\psi(4040) \too \EE)$ and $\Gamma(\psi(4415) \too \EE)$ values from world averages~\cite{pdg}, we can obtain the branching fractions $\mathcal{B}(\psi(4040) \too \eta J/\psi)=(4.2\pm1.2)\times10^{-3}$ or $(4.8\pm1.5)\times10^{-3}$, and $\mathcal{B}(\psi(4415) \too \eta J/\psi)=(1.2\pm0.6)\times10^{-3}$ or $(2.4\pm1.3)\times10^{-3}$. For $Y_{2}$ state, the mass and width are consistent with the state (called $Y(4220)$) found in $\EE \too \omega\chi_{c0}$~\cite{omegachic2}, $\pi^{+}\pi^{-}h_{c}$~\cite{pipihc}, $\pi^{+}\pi^{-}J/\psi$~\cite{pipijpsi}, $\pi^{+}D^{0}D^{*-}$~\cite{piDDstar} and $\pi^{+}\pi^{-}\psi(3686)$~\cite{pipipsip}, so it is reasonable that we take $Y_{2}$ state as $Y(4220)$. Table~\ref{tab:Y4220} lists the parameters for $Y(4220)$ obtained from all decay modes. We use a constant to fit the mass and width of $Y(4220)$, the fit results are shown in Fig.~\ref{fig:fitY4220}. Table~\ref{tab:Y4220} also lists the $Y(4220)$ parameters from fit results, which are $M_{Y(4220)}=(4220.8\pm2.4)$ MeV/$c^{2}$, $\Gamma_{Y(4220)}=(54.8\pm3.3)$ MeV.
\begin{table}[htbp]
\begin{center}
\caption{ The parameters for $Y(4220)$ from different decay modes, and the last line shows the fit results. }
\label{tab:Y4220}
\begin{tabular}{ccc}
  \hline
  \hline
  Decay mode & $M$ (MeV/$c^{2}$) & $\Gamma$ (MeV)  \\
  \hline
  $\omega\chi_{c0}$ & $4226\pm10$ & $39\pm13$ \\
  $\pi^{+}\pi^{-}h_{c}$ & $4218.4^{+5.6}_{-4.6}$ & $66.0^{+12.3}_{-8.3}$ \\
  $\pi^{+}\pi^{-}J/\psi$ & $4222.0\pm3.4$ & $44.1\pm4.8$ \\
  $\pi^{+}D^{0}D^{*-}$ & $4224.8\pm6.9$ & $72.3\pm9.2$ \\
  $\pi^+\pi^- \psi(3686)$ & $4209.1\pm9.8$ & $76.6\pm14.4$ \\
  $\eta J/\psi$ & $4219\pm7$ & $63\pm10$ \\
  Fit results  & $4220.8\pm2.4$ & $54.8\pm3.3$ \\
  \hline
  \hline
\end{tabular}
\end{center}
\end{table}

$Y_{1}$ and $Y_{3}$ states are likely to be $\psi(4040)$ and $\psi(4415)$ states, it is reasonable that $\psi(4040)$ and $\psi(4415)$ can decay to $\eta J/\psi$ because they are considered as $\psi(3S)$ and $\psi(4S)$ states. For $Y(4220)$ state, it also can decay to $\eta J/\psi$, so it is normal that $Y(4220)$ is also considered as $\psi(nS)$ state. In Ref.~\cite{psi4S}, authors suggest $Y(4220)$ is a ``missing $\psi(4S)$" state, which is predicted in Ref.~\cite{psi4S-2,prediction}. Now from the fit results about the cross section of $\eta J/\psi$, the suggestion is reasonable. If we take $Y(4220)$ as $\psi(4S)$ state, then $\psi(4415)$ will be $\psi(5S)$ state. It is very clear for the $\eta J/\psi$ line shape, the three obvious structures are due to $\psi(3S)$, $\psi(4S)$ and $\psi(5S)$ states. If we take the theoretical range $\Gamma(\psi(4S) \too \EE)=0.63\sim0.66$ keV~\cite{psi4S}, and take fit results for $Y_{2}$ from Table~\ref{tab:fitresult1}, we can obtain the branching fraction $\mathcal{B}(\psi(4S) \too \eta J/\psi)=(5.5\pm1.7)\times10^{-3}\sim(10.6\pm2.2)\times10^{-3}$. While Ref.~\cite{psi4S} predicts the upper limit of branching fraction of $\psi(4S) \too \eta J/\psi$ is to be $1.9\times10^{-3}$, so the fit results have some deviations from theory prediction.

Recently, the measurement of $\EE \too \pi^{+}\pi^{-}J/\psi$~\cite{pipijpsi} from BESIII indicates that the previous $Y(4260)$ structure maybe the combination of two resonances, the lower resonance is $Y(4220)$. BESIII also has observed process $\EE \too \gamma X(3872)$~\cite{X3872} around 4.23 and 4.26 GeV. In Ref.~\cite{X3872a,X3872b,X3872c}, authors suggest $X(3872)$ can be identified as a $2^{3}P_{1}$ $\bar{c}c$ state with effect from $\bar{D}D^{*}$ threshold. If we assume $X(3872)$ structure is due to $\chi_{c1}(2P)$ state, and take $Y(4220)$ as $\psi(4S)$ state, the process $\EE \too \gamma X(3872)$~\cite{X3872} around 4.23 and 4.26 GeV observed by BESIII will be a simple transition $\psi(4S) \too \gamma \chi_{c1}(2P)$. We think it is a reasonable explanation for $X(3872)$ and $Y(4220)$. It indicates that maybe some exotic structures are due to the conventional charmonium.

In summary, we fit to the $\EE \too \eta J/\psi$ line shape, three resonant structures are evident. The parameters for the three resonant structures are $M_{1}=(3980\pm17\pm7)$ MeV/$c^{2}$, $\Gamma_{1}=(104\pm32\pm13)$ MeV; $M_{2}=(4219\pm5\pm4)$ MeV/$c^{2}$, $\Gamma_{2}=(63\pm9\pm3)$ MeV; $M_{3}=(4401\pm12\pm4)$ MeV/$c^{2}$, $\Gamma_{3}=(49\pm19\pm4)$ MeV, where the first uncertainties are statistical and the second systematic. We attribute the three structures to $\psi(4040)$, $Y(4220)$ and $\psi(4415)$ states. The branching fractions are $\mathcal{B}(\psi(4040) \too \eta J/\psi)=(4.2\pm1.2)\times10^{-3}$ or $(4.8\pm1.5)\times10^{-3}$, and $\mathcal{B}(\psi(4415) \too \eta J/\psi)=(1.2\pm0.6)\times10^{-3}$ or $(2.4\pm1.3)\times10^{-3}$. We emphasize that the second structure in the $\EE \too \eta J/\psi$ is not from $\psi(4160)$, it is more consistent with $Y(4220)$. Combining all $Y(4220)$ parameters obtained from different decays, we give average parameters for $Y(4220)$, which are $M_{Y(4220)}=(4220.8\pm2.4)$ MeV/$c^{2}$, $\Gamma_{Y(4220)}=(54.8\pm3.3)$ MeV. If we take $Y(4220)$ as $\psi(4S)$ state, then $\psi(4415)$ will be $\psi(5S)$ state. It is very clear for the $\eta J/\psi$ line shape, the three obvious structures are due to $\psi(3S)$, $\psi(4S)$ and $\psi(5S)$ states. If we take the theoretical range $\Gamma(\psi(4S) \too \EE)=0.63\sim0.66$ keV~\cite{psi4S}, we can obtain the branching fraction $\mathcal{B}(\psi(4S) \too \eta J/\psi)=(5.5\pm1.7)\times10^{-3}\sim(10.6\pm2.2)\times10^{-3}$, which has some deviations from theory prediction~\cite{psi4S}. At present, the number of high precision data points in $\EE \too \eta J/\psi$ line shape is relatively small, so more measurements around this energy region are desired, this can be achieved in BESIII and BelleII experiments in the further.

\section*{Acknowledgement}
This work is supported by Nanhu Scholars Program for Young Scholars of Xinyang Normal University and National Natural Science Foundation of China under Contract No.11047145. The data of $\EE \too \eta J/\psi$ cross section supporting this work is from previously reported studies, which has been cited as references~\cite{bes1,bes2,belle}. The authors declare that there is no conflict of interest regarding the publication of this paper.

\end{document}